\documentclass[twocolumn,showpacs,preprintnumbers,amsmath,amssymb]{revtex4}

\usepackage{dcolumn}
\usepackage{bm}
\usepackage{epsfig}%
\usepackage{ulem}%
\usepackage{color}%

\begin{document}


\title{Size and mass of Cooper pairs determined by low-energy $\mu$SR and PNR}

\author{V. Kozhevnikov$^{1}$, A. Suter$^2$, H. Fritzsche$^3$,  V. Gladilin$^4$, A.
Volodin$^5$, J. Cuppens$^5$, T.~Prokscha$^2$, E. Morenzoni$^2$, M.
J. Van Bael$^5$, K. Temst$^6$, C. Van Haesendonck$^5$, J. O.
Indekeu$^7$}

\affiliation{
$^1$Tulsa Community College, Tulsa, Oklahoma 74119, USA \\
$^2$Paul Scherrer Institute, CH-5232 Villigen PSI, Switzerland \\
$^3$Canadian Neutron Beam Centre, National Research Council Canada, Chalk River Laboratories, ON, K0J1J0 Canada\\
$^4$TQC, Universiteit Antwerpen, BE-2610 Antwerpen, Belgium\\
$^5$Laboratory of Solid-State Physics and Magnetism, KU Leuven, BE-3001 Leuven, Belgium\\
$^6$Institute for Nuclear and Radiation Physics, KU Leuven, BE-3001 Leuven, Belgium\\
$^5$Institute for Theoretical Physics, KU Leuven, BE-3001 Leuven, Belgium}. \\


\begin{abstract}
\noindent The Pippard coherence length $\xi_0$ (the size of a Cooper
pair) in an extreme type-I superconductor was determined \textit
{directly} through high-resolution measurement of the nonlocal
electrodynamic effect combining low-energy muon spin rotation
spectroscopy and polarized neutron reflectometry. The
renormalization factor $Z\equiv m_{cp}^*/2m$ ($m_{cp}^*$ and $m$ are
the mass of the Cooper pair and the electron, respectively)
resulting from the electron-phonon interaction, and the temperature
dependent London penetration depth $\lambda_{\rm L}(T)$ were
determined as well. A general expression linking $\xi_0$, $Z$ and
$\lambda_{\rm L}(0)$ is introduced and experimentally verified. This
expression allows one to determine experimentally the Pippard
coherence length in \textit{any} superconductor, independent of
whether the electrodynamics is local or nonlocal, conventional or
unconventional.

\end{abstract}\

\maketitle

A core concept of superconductivity is Cooper pairing of electrons.
Cooper pairs have a characteristic size $\xi_0$, the Pippard
coherence length, and an effective mass $m_{cp}^*\equiv 2Zm$, $m$
being the Coulomb and band-structure effective mass of electron and
$Z$ the renormalization factor, a measure of the electron-boson
coupling strength~\cite{Carbotte}. Only weakly dependent on
temperature $T$, $\xi_0$ provides a reference length for the
fundamental length scales, including in particular the London
penetration depth $\lambda_{\rm L}(T)$, characterizing the decay of
a penetrating magnetic field. In units of $\xi_0$, $\lambda_{\rm
L}(T)$ determines whether the superconductor is described by
\textit{local} or \textit{nonlocal} electrodynamics, and
$\lambda_{\rm L}(0)$ determines whether it is type-I or
type-II~\cite{Tinkham}. To ensure consistency $\xi_0$ and $Z$ should
be measured simultaneously. However, in spite of their importance,
$\xi_0$ and $Z$ have not yet been directly and simultaneously
measured in any superconductor.

As shown by Pippard~\cite{Pippard}, $\xi_0$ can be determined in
\textit{nonlocal} superconductors through measurement of the
nonlocal electrodynamic effect~\cite{IS}. Most superconductors are
local and for those it was stated that $\xi_0$ cannot be measured
directly~\cite{Leggett}. Here we report on a direct and simultaneous
determination of $\xi_0$, $Z$ and $\lambda_{\rm L}$ in a nonlocal
superconductor (In) and answer the question how $\xi_0$ can be
determined experimentally in any superconductor.

When a superconductor is in the Meissner state, an external magnetic
field $B_0$ is completely screened due to a persistent current
running in a thin surface layer over which the field decays to zero.
The layer thickness is of the order of the ``magnetic penetration
depth'' $\lambda\equiv B_0^{-1}\int_0^\infty B(z) dz $, $z$ being
the distance from the surface. If the size of the Cooper pairs is
small ($\xi_0\ll\lambda$), the relationship between the current
density and the vector potential can be treated as local. Then the
field decays as $\exp(-z/\lambda_{\rm L})$, where $\lambda_{\rm L} =
\sqrt {\Lambda c^2 / 4 \pi}$ is the London penetration depth ($c$ is
the speed of light). In the London theory the phenomenological
parameter $\Lambda$ is a function of the mass and the density of
superconducting electrons, none of the two being well
defined~\cite{Tinkham}.

The local approximation is applicable to superconductors with
Ginzburg-Landau parameter $\kappa \gtrsim 1.6$~\cite{Halbritter}.
For superconductors with smaller $\kappa$ the Cooper pair size is no
longer negligible and the current density is determined by the
vector potential averaged over a region of dimension $\xi_0$. The
nonlocality leads to deeper field penetration and to distortion of
the $B(z)$ shape. In the pure limit (elastic mean free path
$\ell\gg\xi_0$) $B(z)$ is a function of both $\lambda_{\rm L}(T)$
and $\xi_0$. Therefore, knowledge of $B(z)$ in low-$\kappa$
materials allows one to determine these two parameters.

As predicted by Pippard and supported by the
Bardeen-Cooper-Schrieffer (BCS) theory, in nonlocal superconductors
$B(z)$ is nonmonotonic with a sign reversal at a certain depth. This
\textit{nonlocal electrodynamic effect} is most pronounced in
extreme type-I superconductors, such as In
($\kappa\thickapprox0.07$); it can also occur in unconventional
superconductors with nodes in the energy gap~\cite{Leggett}.

The nonlocal effect has been directly confirmed by  low-energy muon
spin rotation (LE-$\mu$SR) measurements on Pb, Ta and Nb~\cite{A1,
A2}, and soon after by measurements of polarized neutron
reflectivity (PNR)
 in In~\cite{VK}. In Refs.~\cite{A1,A2} a first attempt
was made to infer $\xi_0$ from $B(z)$ in Pb. However, due to an
unresolved issue of systematic errors only statistical errors were
estimated. One can resolved this issue combining LE-$\mu$SR and PNR
measurements since in the latter the systematic errors can be
excluded.

The knowledge of $B(z)$ in nonlocal superconductors also allows one
to determine $Z$. Electrons near the Fermi surface are dressed by a
cloud of virtual phonons, leading to an enhancement of their
effective mass and consequently to a reduction of the Fermi velocity
$v_{\rm F}$. 
Below the critical temperature $T_c$ the phonon mediated attraction
of elections exceeding their screened Coulomb repulsion results in
formation of the Cooper pairs. In BCS theory electron-electron
coupling is weak: $N(0)V \ll 1$, where $V$ is the pairing potential
and $N(0)$ is the single-spin electron density of states at the
Fermi surface obtained from specific-heat measurements. Eliashberg's
strong-coupling theory (SCT) is free from this limitation and agrees
better with experiments~\cite{Carbotte}.

In SCT the effective mass of electrons near the Fermi surface is
$m^*=Zm$ and consequently the effective mass of the Cooper pairs is
$m_{cp}^*=2Zm$. Correspondingly, $\lambda_{\rm L}$ and $\xi_0$ are
renormalized with respect to their values in the weak-coupling
$(wc)$ limit as
\begin{eqnarray}
\lambda_{\rm L} &=&\sqrt{Z}\ \lambda_{\rm L}^{wc}, \label{eq:lambda_sct} \\
\xi_0 &=& \xi_0^{wc}/Z. \label{eq:xi_sct}
\end{eqnarray}

If $\lambda_{\rm L}(T\rightarrow0)$ is measured, the factor $Z$ can
be inferred from Eq.\,(\ref{eq:lambda_sct}). $\lambda_{\rm
L}^{wc}(0)$ can be obtained from the Faber-Pippard formula for
$\Lambda$ derived for the electrons not interacting with the
lattice~\cite{Faber}. This leads to
\begin{eqnarray}
\lambda_{\rm L}^{wc}(0) = \sqrt{\frac{3c^2}{8\pi e^2 N(0)v_F^2}} \: \: ,
\end{eqnarray}
where $e$ is the electron charge. On the other hand, if $\xi_0$ is
measured, $Z$ can be calculated from Eq.\,(2) using the BCS
definition $\xi_0^{wc}\equiv \hbar v_{\rm F}/\pi\Delta(0)=0.18\hbar
v_{\rm F}/k_B T_c$, where $\Delta(0)\equiv \Delta(T=0)$ is the
energy gap and $h = \hbar\times (2\pi)$ and $k_B$ are the Planck and
Boltzmann constant, respectively.

Unfortunately neither of these approaches is applicable for
quantitative analysis since reliable calculation of $v_{\rm F}$ is
not possible due to the complex topology of the Fermi surface in
polyvalent metals~\cite{Ashcroft}. However, if both $\lambda_{\rm
L}(0)$ and $\xi_0$ are known, one can eliminate $v_{\rm F}$, thus
obtaining
\begin{equation}
Z=\frac{c^2 \hbar^2}{12.5 \pi T_c^2 e^2 \gamma}\cdot\frac{1}{\lambda_{\rm L}(0)^2 \xi_0^2} \: \: ,\label{eq:Z}
\end{equation}

\noindent where it is taken into account that
$N(0)=3\gamma/2\pi^2k_B^2$ ($\gamma$ is the electron heat capacity
coefficient~\cite{Ashcroft}). Eq.\,(\ref{eq:Z}) is a \textit{general
expression} independent of the relationship between the current
density and the vector potential and of the specific nature of the
electron-electron pairing.

In  SCT $Z=1+\lambda_m$, where $\lambda_m$ is a mass-enhancement
parameter calculated in Ref.\,\cite{Carbotte} using tunneling
spectroscopy experimental data. $\lambda_m$ can also be obtained
from McMillan's equation for $T_c$~\cite{Wood}, but this approach is
less reliable~\cite{Carbotte}. Therefore, for nonlocal
superconductors Eq.~(\ref{eq:Z}) furnishes a bridge between the
$B(z)$ and the tunneling data, hence providing an independent test
for the validity of $\lambda_{\rm L}$ and $\xi_0$ inferred from the
$B(z)$ data. On the other hand, the vast majority of superconductors
are local. For those Eq.\,(\ref{eq:Z}) allows one to determine
$\xi_0$ from the measured values of $\lambda_{\rm L}(0)$ and $Z$
using, e.g., LE-$\mu$SR and tunneling experiments, respectively.

\begin{table}
\caption{\label{tab:example} Parameters of the samples.  $d$ is the
thickness, $o$ is the oxide layer thickness (see Ref.~\cite{VK}),
RRR is the residual resistivity ratio; $\ell$ is the elastic mean
free path.}
\begin{ruledtabular}
\begin{tabular}{lllllll}
 sample & $T_c$(K)  & size (cm) & $d$ ($\mu$m) & $o$ (nm) & RRR & $\ell$ ($\mu$m)\\
\hline
    IN-1  &   3.415 &  2$\times$3&  2.5 & $\leq$1 &560& 11 \\
    IN-2  &   3.415 &  ${\O}$ 6    &  3.3 &$\leq$1 & 730& 14 \\
\end{tabular}
\end{ruledtabular}
\end{table}

In this work $B(z)$ is measured using the
LE-$\mu$SR~\cite{Morenzoni, Morenzoni_2} and PNR~\cite{Felcher}
techniques. Combining these complementary techniques enables an
independent verification of the inferred values of $\xi_0$ and
$\lambda_{\rm L}(T)$ and an estimate of there total uncertainties.
The samples were two indium films whose parameters are listed in
table 1. The films were deposited on an oxidized silicon wafer
(IN-1) and on a sapphire crystal (IN-2) held at room temperature via
thermal evaporation of indium shots (99.9999\% purity) at a base
pressure $\lesssim 5\cdot10^{-9} \, {\rm mbar}$. The IN-1 sample was
used for PNR measurements and for LE-$\mu$SR measurements down to
2.9 K. The IN-2 sample was used for LE-$\mu$SR measurements at lower
temperatures.
\begin{figure}
\epsfig{file=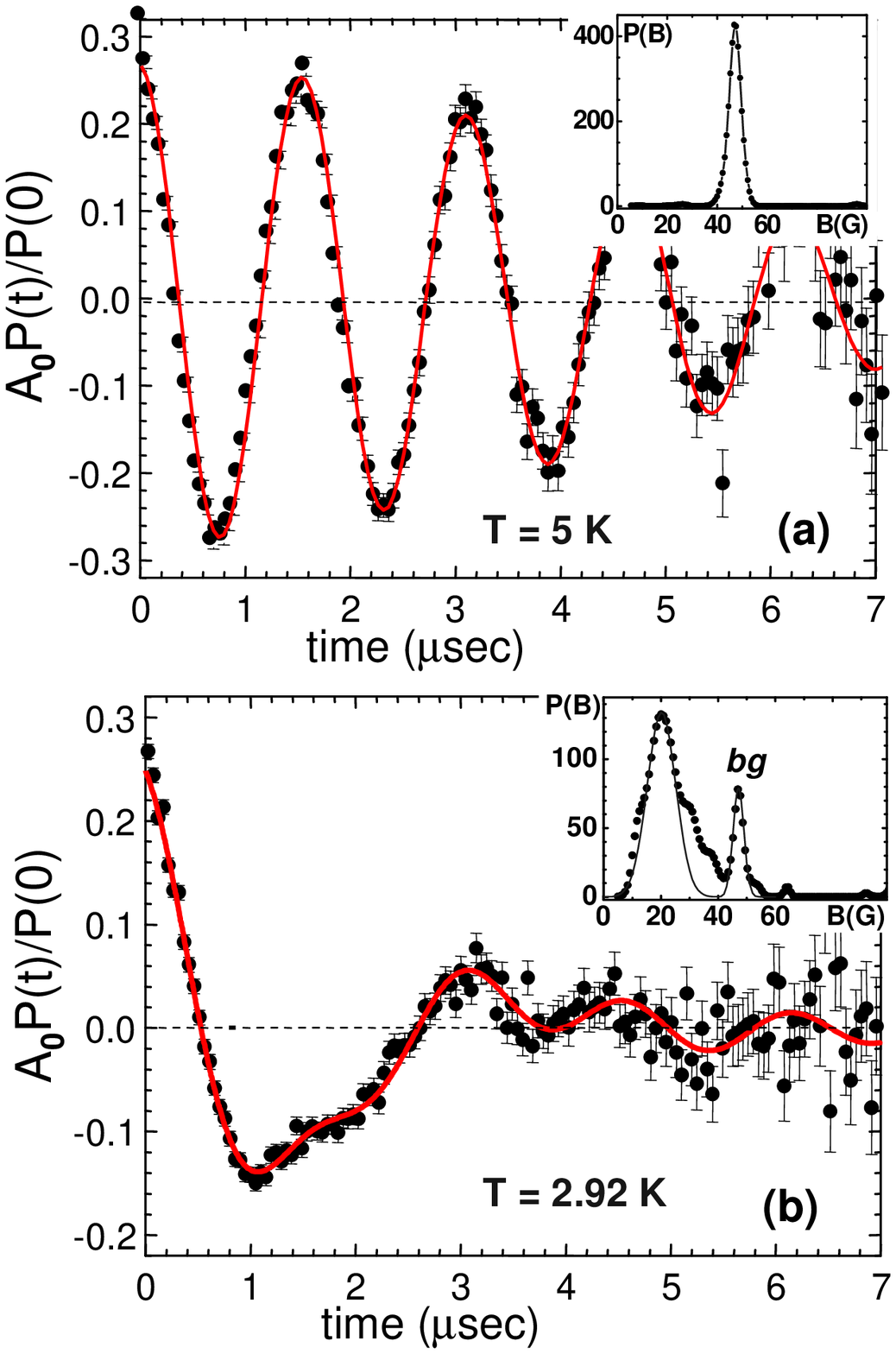,width=6 cm} \caption{\label{fig:epsart} Time
spectra of muons with energy of $22 \, {\rm keV}$ at an applied
field of $47 \, {\rm Oe}$ in the normal state (a) and in the
Meissner state (b). The curves are fits to the Gaussian (a) and the
time-domain (b) models. The inserts present the field spectra
$P(B)$. In the insert in (b) $bg$ is the background contribution,
while the curve represents two Gaussian peaks to approximate the
experimental spectrum. $A_0$ is the maximum observable asymmetry of
the muon decay.}
\end{figure}

The samples' DC magnetization exhibits a clear-cut first-order phase
transition with deep supercooling. The measured critical field
$H_c(T)$ and $T_c$ perfectly agree with values reported in
literature~\cite{Finnemore, Grigoriev}. The surface of the films
consists of nearly atomically flat terraces (root-mean-square
roughness of the terrace areas is $\lesssim 2 \,$\AA) with a typical
size around $5 \, {\rm \mu m}$ with voids in between. This size is
much larger than $\xi_0$ and the total area of the voids does not
exceed 3\% of the sample surface. Therefore the terrace surface
structure should not affect the electrodynamic properties of the
films. For both samples the elastic mean free path $\ell >> \xi_0$,
therefore the samples are type-I superconductors in the pure
limit~\cite{Tinkham}.

The \textbf{LE-$\mu$SR experiments} were performed at the $\mu$E4
beamline of the Swiss Muon Source at the Paul Scherrer
Institute~\cite{Prokscha}. Typical spectra of muon polarization $P$
are presented in Fig.~1. In the normal state $B$ is uniform and the
$P(B)$ spectrum (insert in Fig.~1a) has Gaussian shape due to
nuclear dipole broadening. In the Meissner state $B$ is not uniform
and $P(B)$ is not Gaussian. However, Gaussian spectra (curve in the
insert of Fig.~1b) provide a good first approximation.
\begin{figure}
\epsfig{file=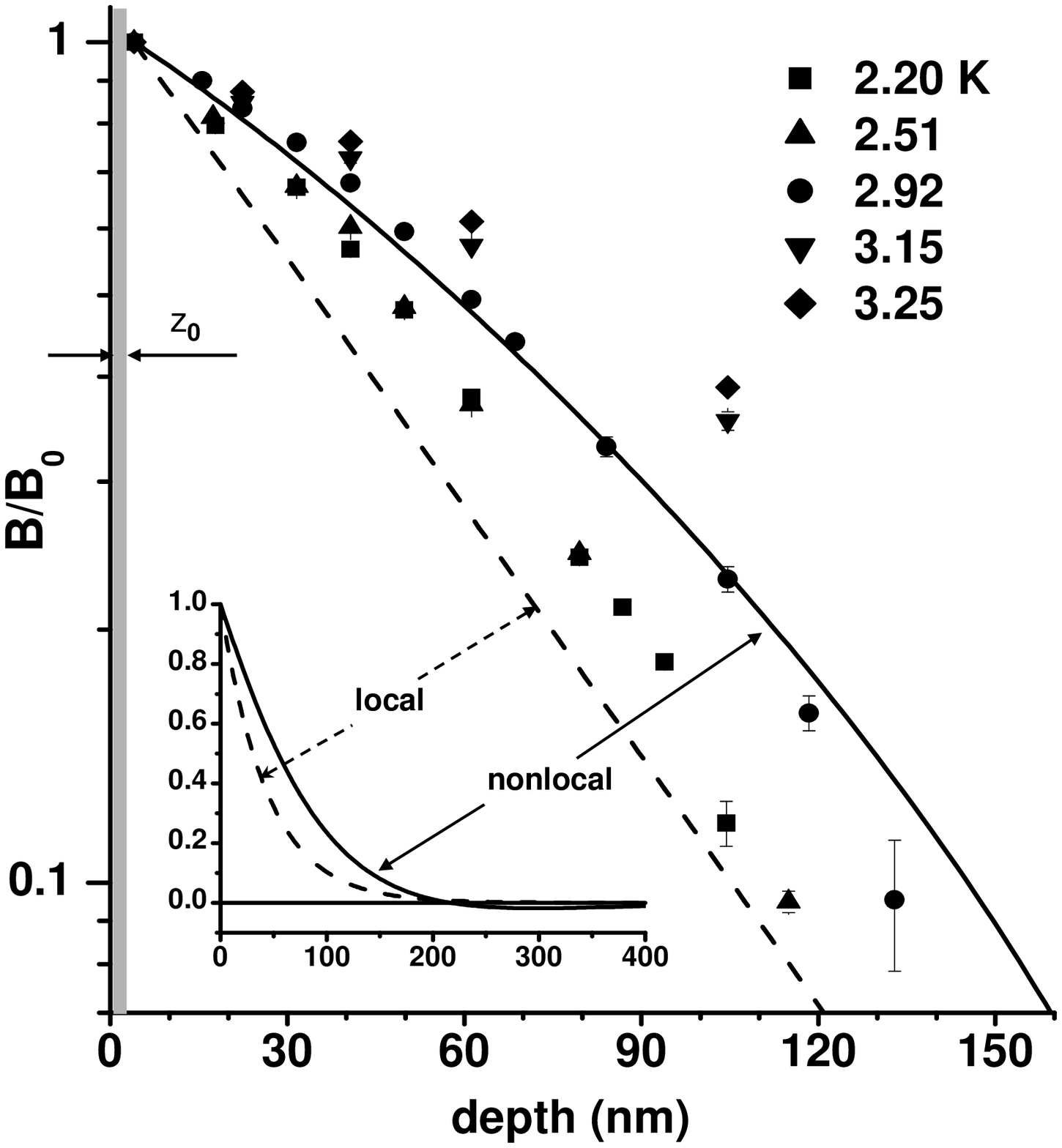,width=6 cm} \caption{\label{fig:epsart}
Reduced field $B/B_0$ obtained from the LE-$\mu$SR measurements
using the Gaussian model. The depth is the average stopping distance
of the implanted muons calculated from the Monte Carlo  TRIM.SP
code. The solid (dashed) line is the field profile calculated from
the nonlocal (local) theory at T = 2.92 K with $\xi_0$=380 nm,
$\lambda_{\rm L}(0)$=30 nm and the dead layer $z_0$= 4 nm, inferred
from the $\chi^2$ and the time-domain analysis. The insert presents
the same field profiles on a linear scale.}
\end{figure}

$B(z)$ points obtained using a Gaussian model \cite{A2} for the
depolarization of the precessing muons are presented in Fig.~2 along
with the $B(z)$ curve calculated from the local and nonlocal
theories . The sample temperature was determined \textit{in situ}
based on the $H_c(T)$ phase diagram obtained from magnetization
measurements. As can be seen in Fig.~2, the ``Gaussian'' $B$-points
exhibit a pronounced non-exponential depth dependence, consistent
with the nonlocal effect. Discrepancies between the points and the
``nonlocal" theoretical curve are mainly caused by incomplete
adequacy of the Gaussian model. On the other hand, the qualitative
consistency of the ``Gaussian'' points with the nonlocal theory
justifies the application of a time-domain model \cite{Kiefl},
directly assuming the ``nonlocal'' shape of $B(z)$ with
$\lambda_{\rm L}$ as an adjustable parameter.
\begin{figure}
\epsfig{file=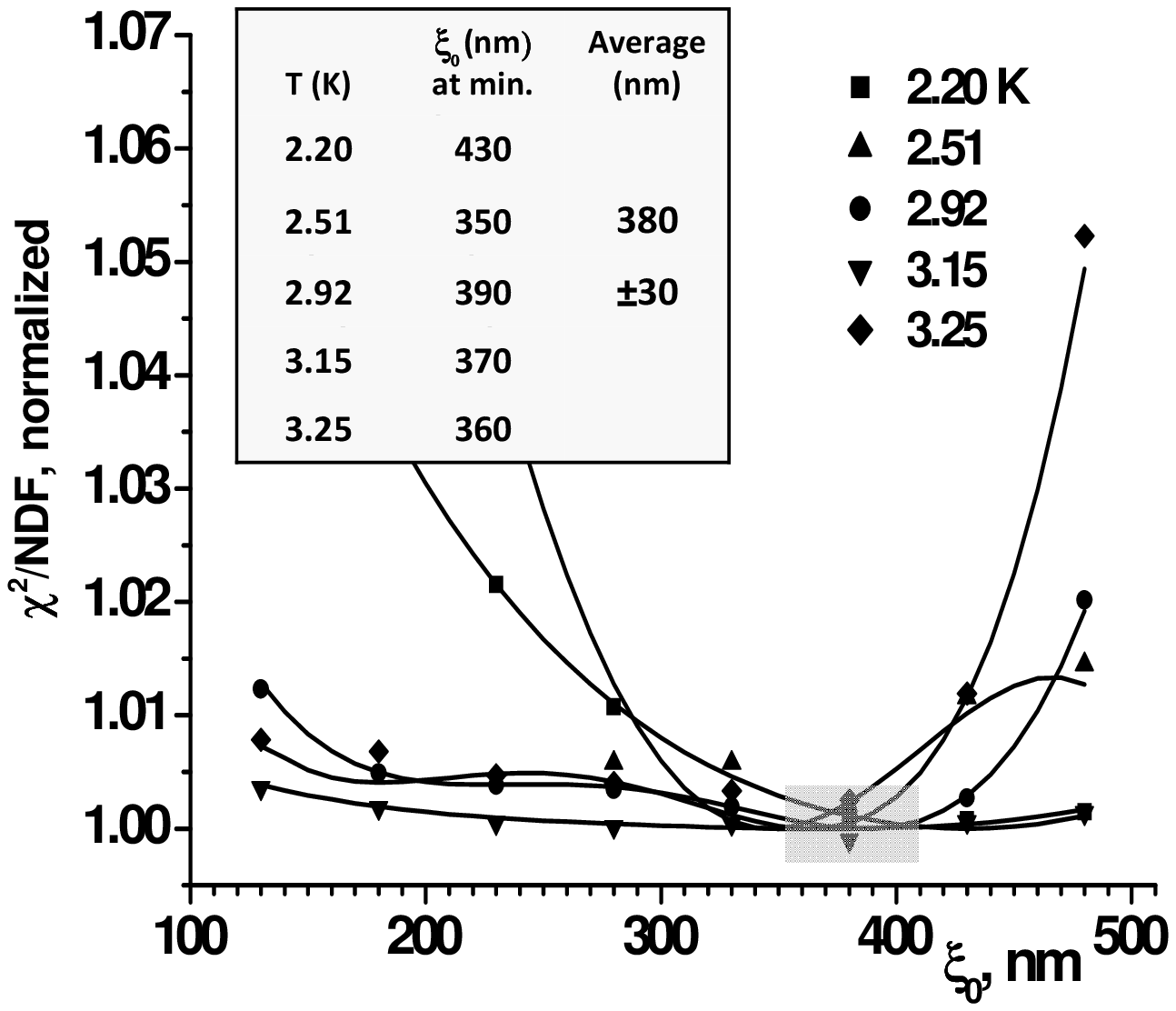,width=6 cm} \caption{\label{fig:epsart}
$\chi^2$/NDF versus $\xi_0$ and the corresponding polynomial
fittings normalized to the minimum value for each temperature. The
shaded area indicates the range of the best estimates for $\xi_0$.
The inserted table gives positions of the minima and their average
with the standard deviation.}
\end{figure}

In nonlocal superconductors the effective penetration depth $\lambda_{\rm eff}\approx(\lambda_{\rm L}^2 \xi_0)^{\frac{1}{3}}$~\cite{Tinkham}. Therefore
either $\lambda_{\rm L}$ or $\xi_0$ has to be determined using an additional criterium. Since $\lambda_{\rm eff}$ is mainly sensitive to $\lambda_{\rm L}$
and $\xi_0$ weakly depends on temperature, this parameter is $\xi_0$. Its optimal value was found from $\chi^2$ analysis of the global fits (simultaneous
fitting of all implantation energy data sets at each temperature). The fits were performed with different $\xi_0$ chosen around a theoretical value of $377
\, {\rm nm}$~\cite{Wood}. The graphs for $\chi^2$/NDF (NDF stands for number of degrees of freedom) versus $\xi_0$ are presented in Fig.~3. Values of
$\xi_0$ at the minima and the best estimate for $\xi_0$ are given in the insert. The values of $\lambda_{\rm L}$ obtained from the global fits with $\xi_0
= 380 \, {\rm nm}$ are presented in a summary graph in Fig.~5.

The \textbf{PNR measurements} were performed on the D3 reflectometer at the NRU reactor in Chalk River. The experimental data and simulations for the
reflectivity of neutrons polarized parallel ($R^+$) and antiparallel ($R^-$) to the magnetic field and for the spin asymmetry $(R^+-R^-)/(R^++R^-)$ are
presented in Fig.~4. In agreement with our LE-$\mu$SR results and previous PNR results~\cite{VK}, the neutron spin asymmetry (Fig.~4, insert (a)) simulated
with $B(z)$ calculated from the nonlocal theory matches the experimental data significantly better than the simulation based on the London field profile.
The ``nonlocal" $B(z)$ was calculated with $\xi_0 = 380 \, {\rm nm}$ obtained from the LE-$\mu$SR data. The best match of the simulated spin asymmetry with
the experimental data was achieved for $\lambda_{\rm L}(T = 0.3 \, \rm{K}) = 28.0 \pm 2.5 \, {\rm nm}$.
\begin{figure}
\epsfig{file=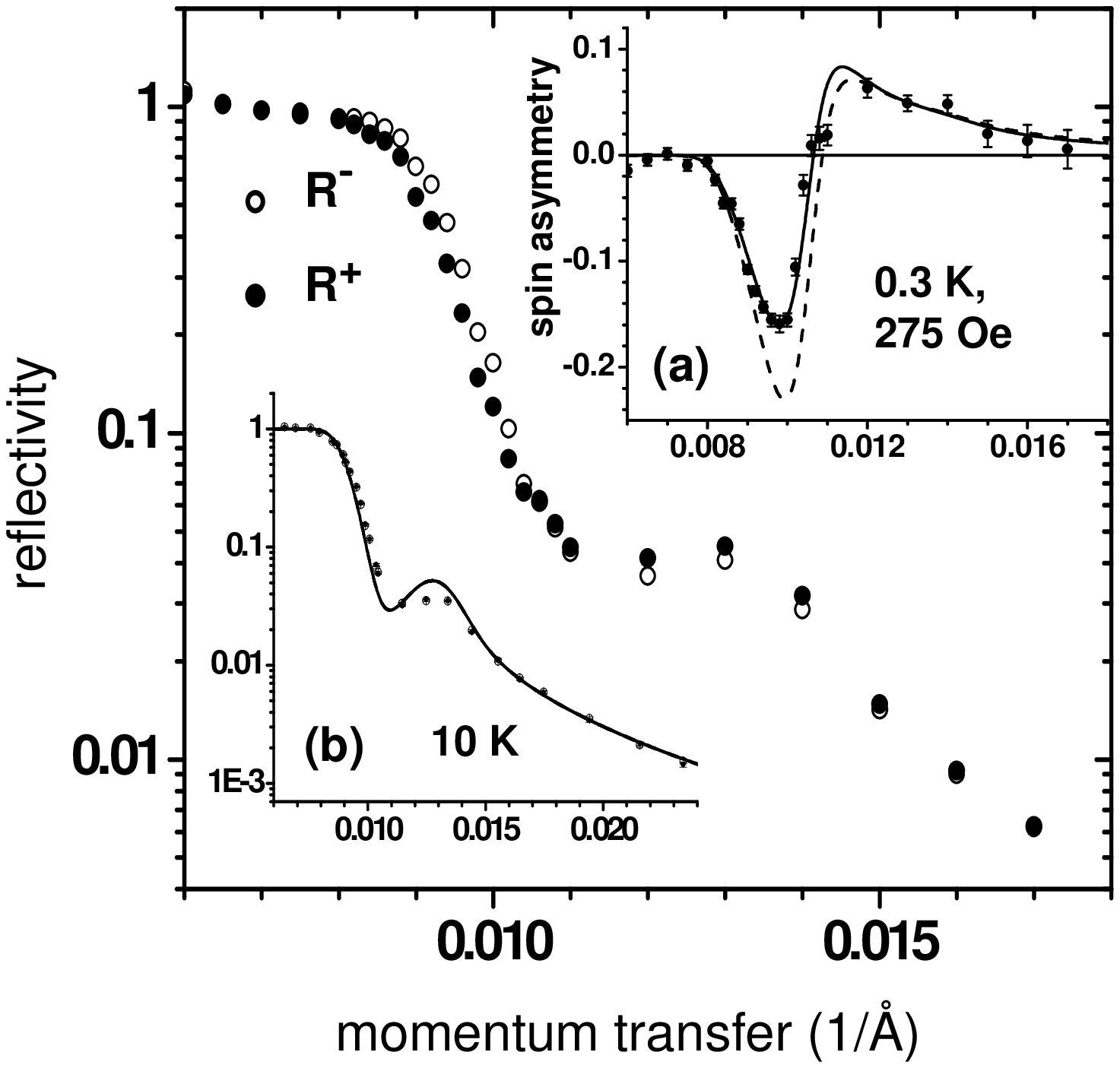,width=6 cm} \caption{\label{fig:epsart}
Neutron reflectivity in the Meissner state at $T = 0.3 \, {\rm K}$
and $B_0 = 275 \, {\rm Oe}$. Insert (a): Measured spin asymmetry
(points) and its simulation with $B(z)$ obtained from the local
(dashed curve) and nonlocal (solid curve) theories with
$\lambda_{\rm L} = 28 \, {\rm nm}$ and $\xi_0 = 380 \, {\rm nm}$.
Insert (b): Measured (points) and simulated (curve) reflectivity in
the normal state.}
\end{figure}

Results for $\lambda_{\rm L}$ inferred from the LE-$\mu$SR and PNR
measurements are shown in Fig.\,5. The values of $\lambda_{\rm
L}(T)$ are consistent with each other and agree with the two-fluid
formula~\cite{Tinkham}. The best estimate of $\lambda_{\rm L}(0)$ is
$30 \pm 2 \, {\rm nm}$.

Having determined $\lambda_{\rm L}(0)$ and $\xi_0$ we calculated $Z$
from Eq.\,(\ref{eq:Z}). The results are summarized in the table
inserted in Fig.\,5, from which it can be concluded that the value
of $Z$ obtained in this work agrees with the value obtained from
electron tunneling data. The latter implies that our results
\textit{quantitatively} confirm the nonlocal electrodynamic effect
and \textit{confirm} the validity of Eq.\,(\ref{eq:Z}).
\begin{figure}
\epsfig{file=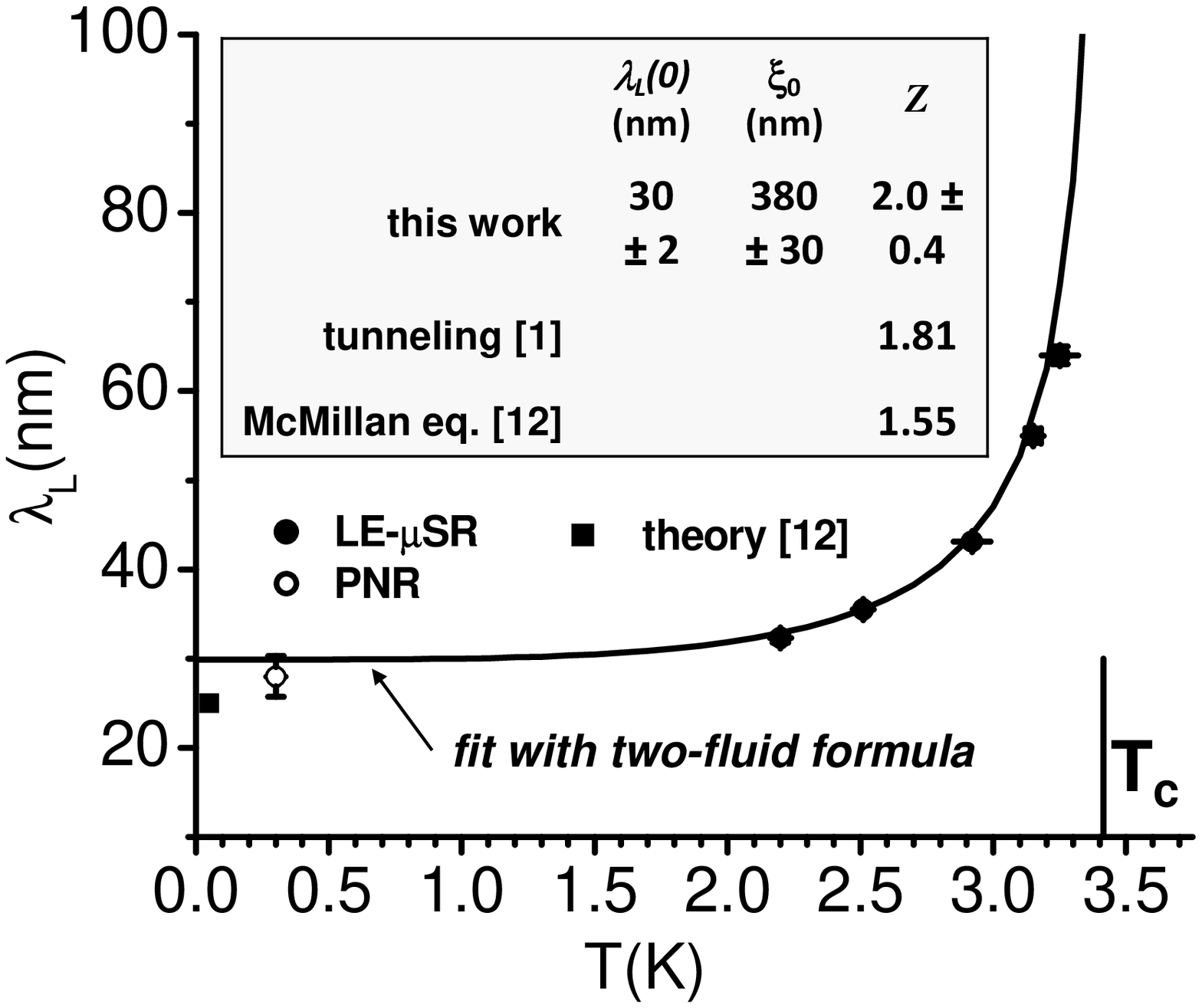,width=6 cm}
\caption{\label{fig:epsart}Temperature dependence of the London
penetration depth in indium.  The inserted table gives the values of
$\lambda_{\rm L}(T\rightarrow0)$, $\xi_0$ and $Z$ obtained in this
work, and the values of $Z$ calculated from the tunneling
spectroscopy data~\cite{Carbotte} and from McMillan's equation for
$T_c$~\cite{Wood}.}
\end{figure}

In conclusion, \textit{high resolution} measurements of the magnetic
field profiles $B(z,T)$ were performed, for the first time, in an
extreme type-I superconductor (indium) combining LE-$\mu$SR and PNR
measurements. The $B(z)$ profiles \textit{quantitatively} confirm
the Pippard/BCS nonlocal electrodynamic effect. The $B(z)$ data were
used to determine the \textit{Pippard coherence length} $\xi_0$, the
\textit{renormalization factor} $Z$ for the electron-phonon mass
enhancement of the Cooper pairs, and the \textit{London penetration
depth} $\lambda_{\rm L}(T)$. The experimentally verified
equation\,(\ref{eq:Z}) allows one to infer $\xi_0$ in \textit{any}
superconductor from the results for $\lambda_{\rm L}(0)$ and $Z$.

\textbf{Acknowledgments}. We thank Tom Moorkens, Maarten Trekels and
Bastian Wojek for help with the measurements. This work was
supported by the National Science Foundation (DMR 0904157) and by
the Science Foundation - Flanders (FWO); J.O.I. acknowledges the
support by KU Leuven grant OT/11/063.


\end{document}